\documentclass[twocolumn,aps,showpacs,floats,prl]{revtex4-1}
\usepackage[usenames,dvipsnames]{color}
\usepackage{amsmath}
\usepackage{amssymb}
\usepackage{graphicx}
\usepackage{anysize}
\usepackage{mathrsfs}
\marginsize{1.5cm}{1.5cm}{0.5cm}{1.cm}
\usepackage[final]{pdfpages}

\usepackage[unicode=true,pdfusetitle, bookmarks=true,bookmarksnumbered=false,bookmarksopen=false, breaklinks=false,pdfborder={0 0 1},backref=false,colorlinks=true]{hyperref}
\hypersetup{colorlinks=true,
	linkcolor=black,
	anchorcolor=brown,
	citecolor=black,
	urlcolor=black
}

\usepackage{breakurl}

\makeatletter
\@ifundefined{textcolor}{}
{
 \definecolor{BLACK}{gray}{0}
 \definecolor{WHITE}{gray}{1}
 \definecolor{RED}{rgb}{1,0,0}
 \definecolor{GREEN}{rgb}{0,1,0}
 \definecolor{BLUE}{rgb}{0,0,1}
 \definecolor{CYAN}{cmyk}{1,0,0,0}
 \definecolor{MAGENTA}{cmyk}{0,1,0,0}
 \definecolor{YELLOW}{cmyk}{0,0,1,0}
}

\usepackage{graphics}
\usepackage{epsfig}
\usepackage{epsf}

\providecommand{\nn}{\nonumber}
\providecommand{\be}{\begin{equation}}
\providecommand{\ee}{\end{equation}}
\providecommand{\bea}{\begin{eqnarray}}
\providecommand{\eea}{\end{eqnarray}}
\providecommand{\beas}{\begin{eqnarray*}}
\providecommand{\eeas}{\end{eqnarray*}}

\providecommand{\beni}{\begin{equation*}}
\providecommand{\eeni}{\end{equation*}}

\providecommand{\bw}{\begin{widetext}}
\providecommand{\ew}{\end{widetext}}

\makeatother

\begin{document}

\title{Inverse spin glass and related maximum entropy problems}

\author{Michele Castellana$^1$ and William Bialek$^{1,2}$}

\affiliation{$^1$Joseph Henry Laboratories of Physics and Lewis--Sigler Institute
for Integrative Genomics, Princeton University, Princeton, New Jersey
08544\\
$^2$Initiative for the Theoretical Sciences, The Graduate Center, City University of New York, 365 Fifth Ave., New York, New York 10016}


\begin{abstract}
If we have a system of binary variables and we measure the pairwise correlations among these variables, then the least structured or maximum entropy model for their joint distribution is an Ising model with pairwise interactions among the spins. Here we consider inhomogeneous systems in which we constrain (for example) not the full matrix of correlations, but only the distribution from which these correlations are drawn.  In this sense, what we have constructed is an inverse spin glass: rather than choosing coupling constants at random from a distribution and calculating correlations, we choose the correlations from a distribution and infer the coupling constants. We argue that such models generate a block structure in the space of couplings, which provides an explicit solution of the inverse problem. This allows us to generate a phase diagram in the space of (measurable) moments of the distribution of correlations. We expect that these ideas will be most useful in building models for systems that are nonequilibrium statistical mechanics problems, such as networks of real neurons.
\end{abstract}

\date{\today}

\maketitle

Systems at thermal equilibrium are in a state of maximum entropy.
But maximizing entropy also provides a method for building models
of systems, whether in equilibrium or not, that are consistent with
some set of measurements but otherwise have as little structure as
possible \cite{jaynes_57}. Concretely, we consider a system described
by variables ${\mathbf{\sigma}}\equiv\{\sigma_{1},\sigma_{2},\cdots,\sigma_{N}\}$,
and we would like to construct the probability distribution $P({\mathbf{\sigma}})$
over these states. We can take from experiment measurements on the
expectation values of various operators $O_{1}({\mathbf{\sigma}}),O_{2}({\mathbf{\sigma}}),\cdots,O_{K}({\mathbf{\sigma}})$,
and so we insist that 
\begin{equation}
\sum_{\mathbf{\sigma}}P({\mathbf{\sigma}})O_{\mu}({\mathbf{\sigma}})=\langle O_{\mu}({\mathbf{\sigma}})\rangle_{{\rm expt}}.\label{constraints1}
\end{equation}
Searching all probability distributions that obey these constraints,
we can find the one which has the maximum entropy, and the result
is a Boltzmann--like distribution $P({\mathbf{\sigma}})=e^{-E({\mathbf{\sigma}})} / Z(\{g_{\mu}\})$, with an effective energy $E({\mathbf{\sigma}})=\sum_{\mu=1}^{K}g_{\mu}O_{\mu}({\mathbf{\sigma}})$,  where $Z(\{g_{\mu}\})$ is the partition function enforcing the normalization of $P(\sigma )$. To complete the construction we must find the values of the coupling
constants $g_{\mu}$ that satisfy the constraints in Eq. (\ref{constraints1}).
This is the inverse of the usual problem in statistical mechanics: rather than knowing the coupling constants and trying
to predict expectation values, we are given the expectation values
and must determine the coupling constants. In general this inverse
problem is hard, and application of the maximum entropy method to
real systems usually depends on detailed numerics.

Recent applications of the maximum entropy approach to a wide variety of biological systems---patterns of activity in networks of neurons \cite{schneidman+al_06,shlens+al_06,tkacik+al_06,yu+al_08,tang+al_08,tkacik+al_09,shlens+al_09,ohiorhenuan+al_10,ganmor+al_11,tkacik+al_12}, the structure and dynamics of biochemical and genetic networks \cite{lezon+al_06,tkacik_07}, the ensemble of amino acid sequences in families of proteins \cite{bialek+ranganathan_07,seno+al_08,weigt+al_09,halabi+al_09,mora+al_10,marks+al_11,sulkowska+al_12}, and ordering in flocks of birds \cite{bialek+al_12,bialek+al_13}---have generated renewed interest in the inverse problem. A variety of approximations and algorithmic solutions have been suggested, based on methods borrowed from statistical mechanics  \cite{cocco+al_11,ricci_12}, statistical inference \cite{decelle_14,kaipio+somersalo_05} and machine learning \cite{cocco+monasson_11}. The essential difficulty is that these systems are strongly inhomogeneous. As an example, if the  $\langle\hat{O}_{\mu}({\mathbf{\sigma}})\rangle_{{\rm expt}}$ are the correlations between the spikes generated by pairs of neurons in a network, in principle we have a correlation matrix with an arbitrary structure and hence $N(N-1)/2$ coupling constants $\{g_{\mu}\}$ that need not have any simple relation to one another. In this setting even the forward problem ($\{g_{\mu}\}\rightarrow\{\langle O_{\mu}({\mathbf{\sigma}})\rangle\}$) is difficult.

One of the lessons from the statistical mechanics of disordered systems is that we can make statements about an ensemble of systems with randomly chosen parameters even if it is difficult to solve the problem of a single system with inhomogeneous parameters \cite{mezard+al_87}. Here we apply this lesson to the inverse problem. Suppose that $\mu$ is a local index referring (for example) to single sites or links in a network: rather than asking for the expectation value of each local operator, we will ask about the distribution of expectation values across the network.
This idea is guided by previous works on maximum entropy models for neural activity, where one considers ensembles of networks constructed by drawing mean spike probabilities and pairwise correlations from the observed distribution of these quantities across a real network, and then solves the full inverse problem for many members of this ensemble \cite{tkacik+al_06,tkacik+al_09}.  Interestingly,  these ``typical'' networks have many properties in common with the real network. The advance here is that the system is insensitive to the precise topology of the network, and the physical information is encoded into the distribution of expectation values of local operators across the network rather than in the expectation values of all local operators: this will be the working hypothesis of the maximum entropy approach presented in this Letter. More concretely, given the moments $M_{n}=\frac{1}{K}\sum_{\mu=1}^{K}\left(\langle O_{\mu}({\mathbf{\sigma}})\rangle_{{\rm expt}}\right)^{n}$ for $n=1,\, 2,\, \cdots ,\, R$,  we  will show an analytic approach to construct the probability distribution over states $P({\mathbf{\sigma}})$ that is consistent with these moments, but otherwise as random as possible.

To maximize the entropy of $P({\mathbf{\sigma}})$ subject to constraints
on the moments $\{M_{n}\}$, we proceed as usual by introducing Lagrange
multipliers, so that we should maximize 
\begin{eqnarray}\label{eq:lagrange_function}
 \mathscr{L} &=& -\sum_{\mathbf{\sigma}}P({\mathbf{\sigma}})\ln P({\mathbf{\sigma}})+N\lambda_{0}\left[\sum_{\mathbf{\sigma}}P({\mathbf{\sigma}})-1\right] +\\ \nonumber 
&  & +N\sum_{n=1}^{R}\lambda_{n}\left[\frac{1}{K}\sum_{\mu=1}^{K}\Bigg(\sum_{\mathbf{\sigma}}P({\mathbf{\sigma}})O_{\mu}({\mathbf{\sigma}})\Bigg)^{n}-M_{n}\right],
\end{eqnarray}
where $\lambda_{0}$ enforces normalization, and we keep the first
$R$ moments. Notice that the first term in Eq. (\ref{eq:lagrange_function})
is extensive; explicit factors of $N$ insure that the other terms also are extensive.  Solving ${\partial{\mathscr{L}}}/{\partial P({\mathbf{\sigma}})}=0$, we find that the maximum entropy distribution is again a Boltzmann distribution, but with the coupling constants related, self--consistently, to the expectation values: 
\be
P({\mathbf{\sigma}})  = \frac{1}{{Z(\{g_{\mu}\})}}\exp\left[\sum_{\mu=1}^{K}g_{\mu}O_{\mu}({\mathbf{\sigma}})\right]\label{gensol1},
\ee
where
$g_{\mu}  \equiv {N\over K} \sum_{n=1}^{R} n\lambda_{n} \phi_{\mu}^{n-1}$, and
$\phi_{\mu}=\sum_{\mathbf{\sigma}}P({\mathbf{\sigma}})O_{\mu}({\mathbf{\sigma}}) $ is the expectation value of $O_{\mu}({\mathbf{\sigma}})$ in the distribution
$P({\mathbf{\sigma}})$; we still must adjust the $\{\lambda_{n}\}$ to match the observed
$\{M_{n}\}$.

In the simplest version of these ideas, the variables $\sigma_{\rm i}$
are Ising spins, and the operators $O_{\mu}(\{\sigma\})=\sigma_{\mu}$
are the individual spins themselves (hence $K=N$). The maximum entropy
model consistent with knowing the expectation values of  every individual spin corresponds to a collection of independent spins in local magnetic fields  
\be
P({\mathbf{\sigma}})=\frac{1}{{Z(\{h_{{\rm i}}\})}}\exp\Bigg(\sum_{{\rm i}=1}^{N}h_{{\rm i}}\sigma_{{\rm i}}\Bigg),\label{eq:independent_spin_p}
\ee
 with $\langle \sigma_{\rm i}\rangle = \tanh h_{\rm i}$, as usual.  What happens if we know only a limited set of moments of the distribution of $\langle \sigma_{\rm i}\rangle$ across the system?  For example, if we know only the first two moments
$m_1 \equiv {1\over N}\sum_{{\rm i}=1}^M \langle \sigma_{\rm i}\rangle, \, m_2 \equiv {1\over N}\sum_{{\rm i}=1}^M \langle \sigma_{\rm i}\rangle^2$, then the definition of $g_\mu$ gives us
 \begin{equation}
h_{\rm i} = \lambda_1 + 2\lambda_2 \langle \sigma_{\rm i}\rangle  
= \lambda_1 + 2\lambda_2\tanh h_{\rm i} .
\label{h2}
\end{equation}
For a given $\lambda_1$, $\lambda_2$, Eq. (\ref{h2}) has only a discrete set of solutions for $h_{\rm i}$. Thus, the maximum entropy model consistent with the mean and variance of the magnetization across an ensemble of spins consists of independent spins in local magnetic fields which can take only discrete values.  As we constrain more and more moments, the analog of Eq. (\ref{h2}) becomes a higher and higher order polynomial in $\tanh h_{\rm i}$, and hence the number of discrete values of $h_{\rm i}$ increases, approaching a continuous distribution in the limit that we know all the moments.

\begin{figure}
\includegraphics[scale=0.6]{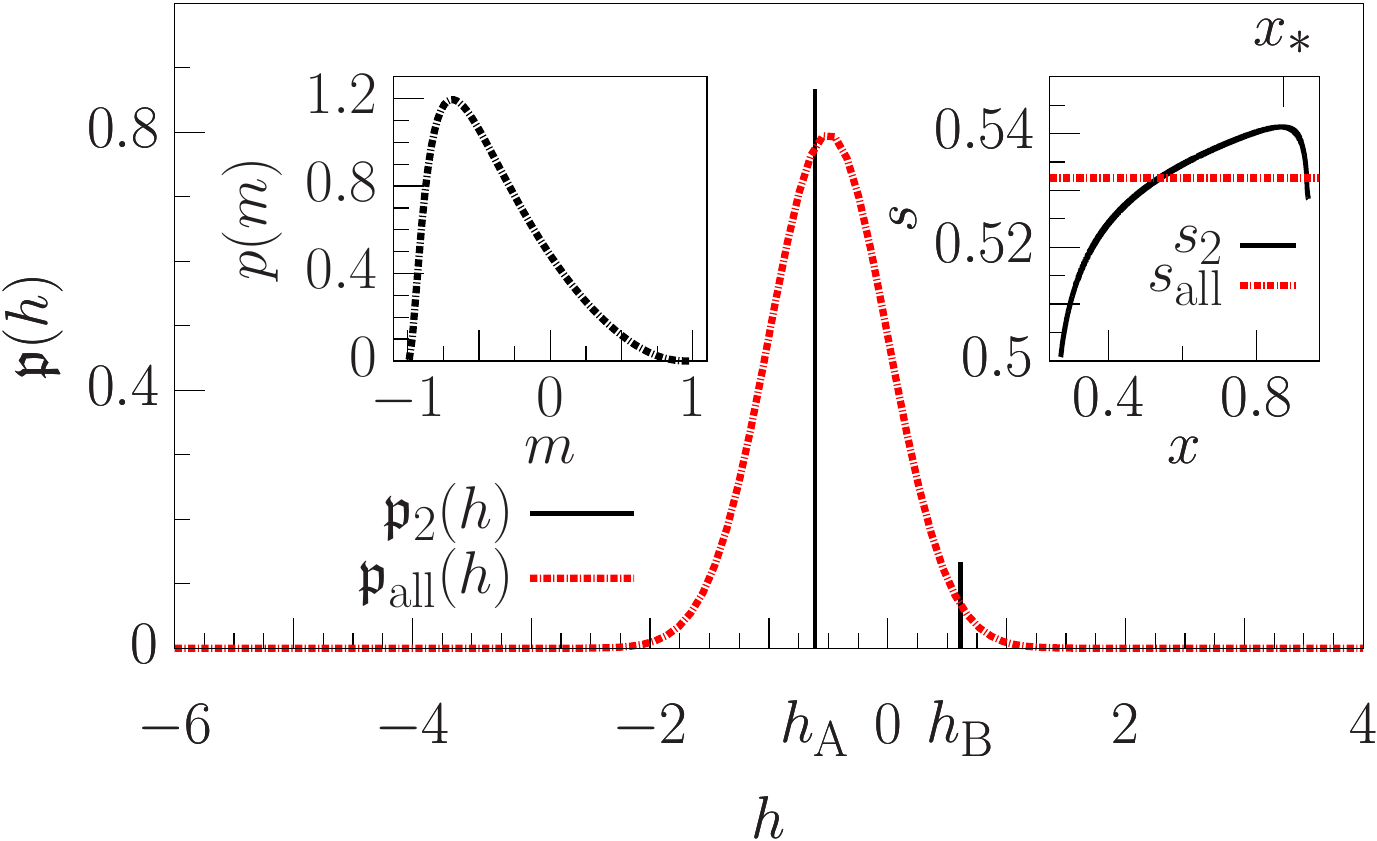}
\caption{
 Maximum entropy for  independent  spins.  The distribution of local  magnetization is shown in the left inset. In the main panel we show the field distributions $\mathfrak{p}_{\textsf{2}}(h)$ (in black)  and $\mathfrak{p}_{\rm{all}}(h)$ (in red) from the maximum entropy solution with two moments and all moments of the local magnetization, respectively. The distribution $\mathfrak{p}_{\textsf{2}}(h)$ is given by two delta peaks at $h=h_{\rm A}$ and $h=h_{\rm B}$.  In the right inset we show the entropy per spin $s_{\textsf{2}}$  (in black) in the two moment case vs. the fraction $x$  of spins in group ${\rm A}$, compared with the entropy in the all moment case $s_{\rm{all}}$ (in red).  The optimal  value $x_{\ast}$ is also marked. \label{fig2}
}
\end{figure}

An illustration of these ideas is shown in Fig. \ref{fig2}.  We choose the expectation values $\langle \sigma_{\rm i}\rangle$ from the distribution shown in the left inset, and build maximum entropy models that are consistent either with knowledge of this full distribution or with just its first two moments.  If the full distribution (all moments) are known, the model has a continuous distribution of fields, and we can compute the resulting maximal entropy, which we denote by  $s_{\rm all}$ (see the Supplemental Material).  Fixing just the first two moments, we assume that there are two groups of spins ${\rm A}$ and ${\rm B}$, with two discrete values of the field $h_{\rm A}$ and $h_{\rm B}$ acting on each group, and thus two values $m_{\rm A}$, $m_{\rm B}$ of the local magnetizations $\langle \sigma_i \rangle$.  Given a fraction $x$ of spins in group ${\rm A}$, we determine $m_{\rm A}$, $m_{\rm B}$ by matching the first two moments
\bea \label{mImIIind}
m_1  =  x\,  m_{\rm A} + (1-x) m_{\rm B}, \; \; \; m_2  =  x\,  m_{\rm A}^2 + (1-x) m_{\rm B}^2,\;\;\;\;
\eea
we plug the solution into the equation $\langle \sigma_{\rm i}\rangle = \tanh h_{\rm i}$, and we solve for $\lambda_1$, $\lambda_2$ by using  Eq. (\ref{h2}): as a result, the entropy $s_{\rm 2}$ depends only on the spin fraction $x$, and we fix $x$ by maximizing $s_{\rm 2}$. It can be shown that this two block ansatz is exact. Indeed, we can fix $\lambda_1$, $\lambda_2$ so that there are three distinct solutions, and the entropy depends on $\lambda_1, \lambda_2$ only: we then maximize the entropy as a function of $\lambda_1$, $\lambda_2$, and at the maximum the fraction of spins $\sigma_{\rm i}$ with local field $h_{\rm i}$ equal to the third solution is equal to zero, and we are left with two values of the local fields (see the Supplemental Material). Importantly,  Fig. \ref{fig2} shows that a weakly constrained, random choice of magnetizations leads to a highly structured bimodal distribution of fields, even though we maximize the entropy and thus minimize structure in the distribution of spin configurations. Given that magnetizations and magnetic fields are related by the identity $\langle \sigma_{\rm i} \rangle = \tanh h_{\rm i}$, if the width of the magnetization distribution is increased---i.e. the standard deviation gets much larger than the mean---the distribution of fields tends to a bimodal distribution composed of two distant peaks rather than to a smooth distribution with a large width. \\

Maximum entropy models are much richer when the operators $O_{\mu}$ live on the links between elements in a network rather than on nodes. Let us consider, then, the case where  $\mu = ({\rm i},{\rm j})$ denotes a pair of  spins, and the operators $O_{\mu}(\{\sigma\})\equiv \sigma_{\rm i} \sigma_{\rm j}$.  The number of operators $K$ is the number of distinct pairs, $N_p =N(N-1)/2$, and we write
$C_{\rm ij}=\langle \sigma_{\rm i}\sigma_{\rm j} \rangle$.  In what follows,  we constrain the first two moments of the correlation distribution,
$C_{1}  =  \frac{1}{{N_{p}}}\sum_{{\rm i}> {\rm j}}C_{{\rm ij}},\; C_{2}  =  \frac{1}{{N_{p}}}\sum_{{\rm i}> {\rm j}}C_{{\rm ij}}^{2}$.
Equation (\ref{gensol1}) then becomes
\begin{eqnarray}\label{eqP}
P(\sigma) &=& \frac{1}{Z} \exp\bigg( \frac{N }{ 2 N_p}  \sum_{{\rm i} ,{\rm j}} J_{\rm ij} \sigma_{\rm i} \sigma_{\rm j}  \bigg), \label{P1}\\
J_{\rm ij} &= & \lambda_1  +  2 \lambda_2 C_{\rm ij}, \label{selfconJ}
\end{eqnarray} 
where in Eq. (\ref{P1}) we have incorporated a diagonal term with $\rm i=\rm j$, which is independent of $\sigma$, and the Lagrange multipliers are set by matching the moments of expectation values
\be  \label{Ca}
C_1  =   -\frac{\partial f}{\partial \lambda_1} - \frac{N}{2N_p}, \; C_2 =  - {1\over 2} \frac{\partial f}{\partial \lambda_2} -  \frac{N}{2N_p} C_{\textrm{I}},
\ee
where the free energy per spin is $f = -(\ln Z) /N$. 
Thus, the system is an Ising model in which the spin--spin couplings $J_{\rm ij}$ are related, bond by bond, to the spin--spin correlations $C_{\rm ij}$. As with Eq. (\ref{h2}), it is difficult to imagine how the self--consistency condition in Eq. (\ref{selfconJ}) can be satisfied by a broad distribution of couplings $J_{\rm ij}$.  In a system that is highly interconnected, the correlations between any pair of spins are dominated by the presence of multiple indirect paths, so that the $C_{\rm ij}$ need not even be correlated with the corresponding direct interactions $J_{\rm ij}$. How then can Eq. (\ref{selfconJ}) be solved?  As with the case of independent spins, we suspect that the self--consistency condition in Eq. (\ref{selfconJ}) can be satisfied only if the system breaks into blocks.  If there are only a discrete set of possible $J_{\rm ij}$, it seems possible that there will be only a discrete set of $C_{\rm ij}$, and that we can arrange the pairs so that $C_{\rm ij}$ and $J_{\rm ij}$ are related linearly.  With $\lambda_1$ and $\lambda_2$ fixed, we have done numerical experiments on systems of up to $N=16$  spins, solving Eq. (\ref{selfconJ}) for  the variables $\{ J_{\rm ij} \}$, and we have found that the couplings $J_{\rm ij}$ are driven to consist of two discrete values (see the Supplemental Material).

Guided by  our numerical experiments, and by the case of independent spins above, we try a block ansatz: we divide the spins $\{\sigma_{\rm i}\}$ into two blocks, ${\rm A} \equiv \{ \sigma_1, \cdots, \sigma_{N_{\rm A}} \}$, and ${\rm B} \equiv \{ \sigma_{N_{\rm A}+1} , \cdots, \sigma_N\}$, and we assume that correlations between spins within a block take the value $C_{\rm I}$ while correlations between spins in different blocks take the value $C_{\rm II}$; $x= N_{{\rm A}}/N$ is the fraction of spins in block ${\rm A}$. The parameters  $C_{\textrm{I}}, C_{\textrm{II}}$ are related to the moments by
\bea 
C_1 & = & [x^2 + (1-x)^2] C_{\textrm{I}} + 2 x (1-x) C_{\textrm{II}} ,\label{Cb}\\ \nonumber 
C_2 & = & [x^2 + (1-x)^2] C_{\textrm{I}} ^2+ 2 x (1-x) C_{\textrm{II}}^2.
\eea
The value of $x$ will be set at the end of the calculation by maximizing the entropy, as  above. It can be shown (see the Supplemental Material) that if the correlations satisfy the high temperature scaling  $C_1 = {\cal O}(1/N)$, $C_2 = {\cal O}(1/N^2)$, the entropy per spin is 
\be\label{s}
s = \ln 2 - \frac{1}{N} \left[ {1\over{2}}\ln {\mathscr{D}}  + \lambda_1 (NC_1 + 1) + 2\lambda_2 (NC_2 + C_{\textrm{I}}) \right],
\ee
where ${\mathscr{D}} = \det\left( 1-2 \, \textrm{diag}(x,1-x)^{-1} \cdot \mathscr{M} \right)$, $\textrm{diag}(x,1-x)$ is a $2 \times 2$ diagonal matrix with diagonal entries $x$, $1-x$, and 
\be\label{eqM}
{\mathscr{M}} = \left(
\begin{array}{cc}
 x^2 (\lambda_1   + 2 \lambda_2 C_{\rm I}) & x(1-x) (\lambda_1 + 2  \lambda_2 C_{\rm II}) \vspace{0.2cm}\\
x(1-x)  (\lambda_1 +2 \lambda_2 C_{\rm II}) & (1-x)^2 ( \lambda_1  + 2 \lambda_2 C_{\rm I} )
\end{array}
\right). 
\ee

\begin{figure}
\includegraphics[scale=0.38]{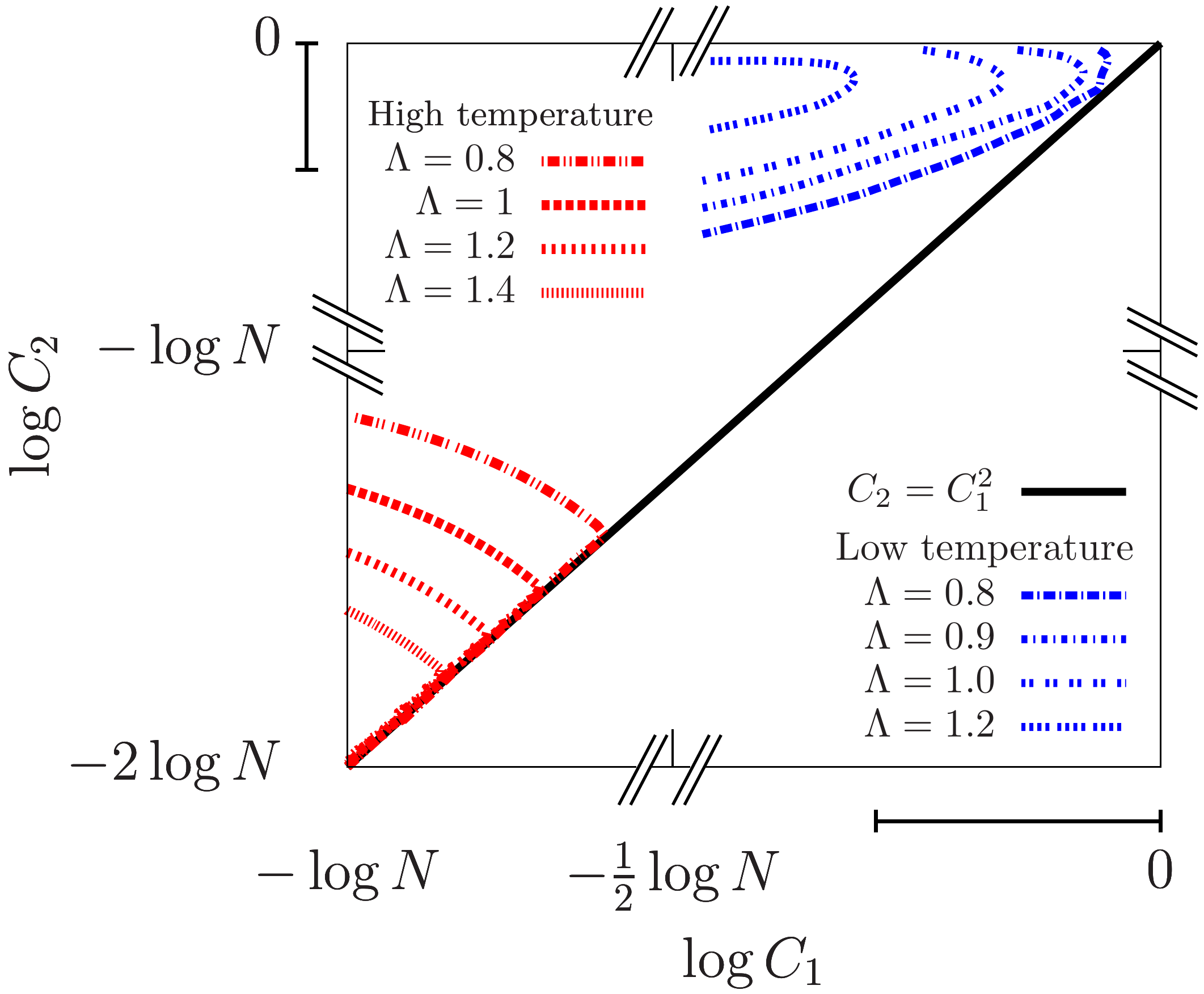}
\caption{Contour plot of the smallest eigenvalue $\Lambda$ of the free energy Hessian as a function of  $\ln C_1, \ln C_2$  for the correlated spin case in the allowed region $C_2 \geq C_1^2$. There is a high temperature phase (in red)  where $C_1 = {\cal O}(1/N)$, $C_2 = {\cal O}(1/N^2)$, a low-temperature phase (in blue) where  $C_1 = {\cal O}(1)$, $C_2 = {\cal O}(1)$, and a critical regime where $C_1 = {\cal O}(1/\sqrt{N})$, $C_2 = {\cal O}(1/N)$. Scale bars on the $x$ and $y$ axis represent one unit of $\ln C_1$ and $\ln C_2$ respectively. In the high temperature phase and in the critical regime, $\Lambda$ is a function of the scaled correlations $N C_1$, $N^2 C_2$ and $\sqrt{N} C_1$, $N C_2$ respectively.  \label{fig1}}
\end{figure}

 Thus, at fixed $x$ we can solve Eqs. (\ref{Ca},\ref{Cb}), and then we can find the value of $x$ that maximizes $s$.  Along the same lines, one can solve the maximum entropy problem in the low temperature phase where $C_1 = {\cal O}(1)$, $C_2 = {\cal O}(1)$.
We can thus draw a phase diagram directly in the space of the observed moments $C_1$ and $C_2$, which we plot as follows:  we represent the partition function as an integral over two order parameters $(m_{\rm A}$, $m_{\rm B}) \equiv \vec{m}$, i.e. $Z \propto \int d \vec{m}  \exp[-N {\cal S}(\vec{m})]$, where
\bea\label{eqS}\nn
{\cal S}(\vec{m})& =& \frac{1}{2} \vec{m}^{\textrm{T}} \cdot {\mathscr{M}}^{-1} \cdot \vec{m} - x \log 2 \cosh(\sqrt{2} m_{\rm A}/x)+ \\
&&- (1-x) \log 2 \cosh (\sqrt{2} m_{\rm B}/(1-x)). 
\eea
Then, we consider the smallest Hessian eigenvalue $\Lambda$ of $\cal S$ computed at the saddle point $\vec{m}^{\ast}$ of the integral in $Z$, and   in Fig. \ref{fig1}  we show contour plots of $\Lambda$ evaluated at the solution of the maximum entropy problem. There is a high temperature phase $C_1 = {\cal O}(1/N)$, $C_2 = {\cal O}(1/N^2)$, and a low temperature phase $C_1 = {\cal O}(1)$, $C_2 = {\cal O}(1)$: in both these phases $\Lambda = {\cal O}(1)$. The high and low temperature phases are separated by a critical regime $C_1 = {\cal O}(1/\sqrt{N})$, $C_2 = {\cal O}(1/N)$ where $\Lambda = {\cal O}(1/\sqrt{N})$. We see that $\Lambda$ gets small close to the boundary of the allowed values $C_2 = C_1^2$: the contours of constant $\Lambda$ bend backward, however, suggesting that we can reach a critical regime $\Lambda = {\cal O}(1/\sqrt{N})$ if  $C_2 = {\cal O}(1/ N)$ and $C_1 = {\cal O}(1/N)$, and we have verified this analytically.

How does  Fig. \ref{fig1} relate to what we know about the phase diagrams of Ising models?  For a ferromagnetic Ising model on a $d$-dimensional hypercube with nearest-neighbor couplings $J_{\rm ij}  = J >0$, the correlations at the critical point are $C_1 \sim \sqrt{C_2} \sim 1/N^\omega$, with $\omega = (d-2+\eta)/d$; we have $\omega = 1/8, 0.3455 \ldots, 1/2$ for $d = 2,3,4$, respectively.  Keeping just two moments our maximum entropy model matches the mean field critical behavior expected at $d=4$. For an  Ising model on a $d$-dimensional hypercube with antiferromagnetic nearest-neighbor interactions $J_{\rm ij} = J < 0$, the lattice can be divided into two embedded sublattices with spins up and down respectively: as a result, roughly half of the spin pairs are positively correlated and the other half are negatively correlated, so $C_1 \sim 1/N$, and in the critical regime  $C_2 \sim 1/N^{2 \omega} \gg C_1^2$: This is again in  reasonably good agreement with the model prediction that there is a critical behavior for $C_1 = {\cal O}(1/N)$, $C_2 = {\cal O}( 1/N) $.  
We note that both in  the one moment and two moment case our maximum entropy solution provides a mean field critical scaling $C_1 \sim \sqrt{C_2} \sim 1/N^{1/2}$, where the fractional exponent $1/2$ results from a Taylor expansion of the free energy in the neighborhood of the high-temperature saddle point $\vec{m}^{\ast} = \vec{0}$. In order to obtain a non-mean-field exponent, one needs to consider an infinite number of moments, so that the free energy can become a non-analytic function of $\vec{m}$ at the critical point.
To conclude, the analysis  for the case where we fix just two moments in the distribution of pairwise correlations seems (barely) sufficient to identify critical behavior in simple ferromagnets and antiferromagnets.  In addition, by fixing at least three moments of the correlations, one obtains a maximum entropy solution with frustrated spin-spin interactions which describes a spin glass (see Supplemental Material).

\textit{Discussion} --  The maximum entropy method is an approach to build models for the joint probability distribution of many degrees of freedom, making use only of measured  expectation values for a limited number of operators.  This approach is attractive both because it picks out the least structured model consistent with the data and because this model is exactly a statistical mechanics problem---the Boltzmann distribution with an energy landscape composed of a sum of terms, one for each of the measured expectation values.  As noted at the outset, the maximum entropy construction can thus be thought of as an inverse statistical mechanics problem, mapping expectation values back to the coupling constants in the effective Hamiltonian.  In this work we have used the maximum entropy method, but in looking at a strongly inhomogeneous system we have constrained not the expectation values of each local operator, but rather the distribution of these expectation values across the (large) system.  In this sense, what we have constructed is an inverse spin glass: rather than choosing coupling constants at random from a distribution and calculating correlations, we choose the correlations from a distribution and infer the coupling constants.

In the Sherrington--Kirkpatrick spin glass, complete randomness in the coupling constants drives the emergence a rich, ordered structure in the pattern of correlations among spins \cite{mezard+al_87}.  Here we have seen that knowing only the distribution of correlations leads to surprising structure in the coupling constants: the values of the spin couplings break into blocks, and with this ansatz we can derive a phase diagram in the space of moments.  Although this needs to be checked carefully, it is natural to conjecture that the number of blocks grows with the number of moments that we constrain.

\begin{acknowledgments}
We thank A Cavagna, P Del Giudice, I Giardina, E Marinari, T Mora, G Parisi, and G Tka\v{c}ik for helpful discussions. WB is especially grateful to his colleagues
in Rome for their hospitality on several visits that influenced this
work. Research supported in part by NSF Grants PHY--0957573, PHY-1305525 and CCF--0939370, by the Human Frontiers Science Program, by
the Swartz Foundation, and by the WM Keck Foundation.
\end{acknowledgments}

\bibliographystyle{unsrtnat}

\clearpage


\section*{Supplementary material}

# Supplemental Material for "Inverse spin glass and related maximum entropy problems"

Michele Castellana[1] and William Bialek[1,2]

[1]*Joseph Henry Laboratories of Physics and Lewis–Sigler Institute for Integrative Genomics,*
*Princeton University, Princeton, New Jersey 08544*
[2]*Initiative for the Theoretical Sciences, The Graduate Center,*
*City University of New York, 365 Fifth Ave., New York, New York 10016*
(Dated: September 7, 2014)

## INDEPENDENT SPINS

Here we show how to obtain the maximum entropy solution shown in Fig. 1 where we constrain all moments of the magnetization in the independent spin case. The Lagrange function (2) reads

$$\mathscr{L} = -\sum_\sigma P(\sigma)\ln P(\sigma) + N\lambda_0\left[\sum_\sigma P(\sigma) - 1\right] \quad \text{(S1)}$$
$$+ N\int dm\,\lambda(m)\left[\frac{1}{N}\sum_{i=1}^N \delta(\langle\sigma_i\rangle - m) - p(m)\right],$$

where $\{\lambda(m)\}_m$ is a continuum set of Lagrange multipliers enforcing the constraint of the magnetization distribution $p(m)$ for every $m$. By taking $\partial\mathscr{L}/\partial P(\sigma) = 0$ we have

$$P(\sigma) = \frac{1}{Z}\exp\left(\sum_{i=1}^N h_i\sigma_i\right), \quad \text{(S2)}$$

where

$$h_i = -\left.\frac{d\lambda(m)}{dm}\right|_{m=\langle\sigma_i\rangle}. \quad \text{(S3)}$$

The magnetic fields $h_i$ are related to the local magnetizations $\langle\sigma_i\rangle$ by the equation

$$\langle\sigma_i\rangle = \tanh h_i. \quad \text{(S4)}$$

From Eq. (S4) we can compute the distribution of fields

$$\mathfrak{p}_{\text{all}}(h) = \frac{1}{N}\sum_{i=1}^N \delta(h_i - h) \quad \text{(S5)}$$
$$= p(\tanh(h))[1 - \tanh(h)^2].$$

The entropy per spin is given by

$$s = -\frac{1}{N}\sum_\sigma P(\sigma)\ln P(\sigma), \quad \text{(S6)}$$

and from Eq. (S5) we obtain the entropy in the all-moment case

$$s_{\text{all}} = \int dh\,\mathfrak{p}_{\text{all}}(h)[\ln(2\cosh h) - h\tanh h]. \quad \text{(S7)}$$

If we constrain only the first two moments, it can be shown that the magnetic fields $h_i$ can take only two values: indeed, let us fix $\lambda_1, \lambda_2$ such that there are three solutions for $\langle\sigma_i\rangle$ to Eq. (5), that we call $m_A, m_B, m_C$. Let us denote by $x_A, x_B, x_C$, with $x_B = 1 - (x_A + x_C)$, the fraction of spins with $\langle\sigma_i\rangle$ equal to $m_A, m_B, m_C$ respectively: we determine $x_A, x_C$ by solving the two moment constraint

$$m_1 = x_A m_A + (1 - (x_A + x_C))m_B + x_C m_C,$$
$$m_2 = x_A m_A^2 + (1 - (x_A + x_C))m_B^2 + x_C m_C^2,$$

and the entropy $s$ now depends only on $\lambda_1, \lambda_2$. If we now maximize $s$ with respect to $\lambda_1, \lambda_2$, the maximal entropy is obtained for $x_C = 0$, and it coincides with the one obtained with our two block ansatz where the magnetic fields take only two values $h_A, h_B$. Along the same lines, if we constrain more than two moments, the correct number of distinct solutions $h_A, h_B, \ldots$ of the analog of Eq. (5) can be singled out as the solutions providing the largest entropy.

## COUPLED SPINS

### One moment case

Here we will discuss briefly the maximum entropy solution where only the first correlation moment is constrained: more details will be given in the calculation for the two moment case. If we constrain only $C_1$, then Eq. (7) becomes the familiar mean–field ferromagnet: the partition function $Z$ can be written as an integral over the order parameter, i.e. $Z \propto \int dm\exp[-N\mathcal{S}(m)]$. So long as $NC_1$ is of order unity the system is in its high temperature phase. If $NC_1$ is of order $\sqrt{N}$, however, $\lambda_1$ is driven close enough to its critical value that we must go to higher order in the expansion around the saddle point $m^*$ of the integral in $Z$. When $\sqrt{N}C_1 > 1.17083\ldots$, there is no consistent solution in the high temperature phase: the only way to match the observed $C_1$ is if this mean correlation has a contribution from a nonzero mean magnetization, and the system is in its low temperature phase. In order to draw the phase diagram of the system, we consider the second derivative of $\mathcal{S}(m)$ at the saddle point $\Lambda \equiv d^2\mathcal{S}/dm^2\big|_{m^*}$. This quantity indicates the stability of the saddle point: a small value of $\Lambda$ implies that the saddle point is nearly unstable, and thus the system

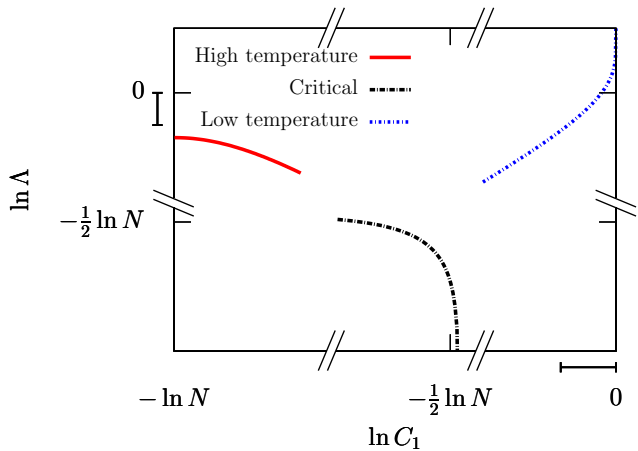

FIG. S1: Phase diagram for coupled spins in the one moment case: second derivative $\Lambda$ of $\mathcal{S}(m)$ at the saddle point $m = m^*$ as a function of $C_1$ in the high temperature phase where $C_1 = \mathcal{O}(1/N)$ and $\Lambda = \mathcal{O}(1)$ (in red), in the critical region where $C_1 = \mathcal{O}(1/\sqrt{N})$ and $\Lambda = \mathcal{O}(1/\sqrt{N})$ (in black), and in the low temperature phase where $C_1 = \mathcal{O}(1)$ and $\Lambda = \mathcal{O}(1)$ (in blue). The scale bars on the $x$ and $y$ axes represent one unit of $\ln C_1$ and $\ln \Lambda$ respectively.

is close to a critical point. In Fig. S1 we plot $\Lambda$ vs. $C_1$ in the high and low temperature phase and in the critical region: thus, even in the problem where we constrain only one moment of the distribution of correlations, we can draw a phase diagram along the $C_1$ axis, without reference to any other parameters.

### Two moment case

First, we performed numerical experiments to verify that Eq. (8) has only two solutions: we fixed $\lambda_1, \lambda_2$ and we solved Eq. (8) for the variables $\{J_{ij}\}$. The solution of these equations involves the exact computation of the correlation functions $\langle \sigma_i \sigma_j \rangle$, which involves $2^N$ terms: hence, this numerical experiment is limited to small values of $N$. An example of this computation is shown in Fig S2: for $N=4$ the distribution of coupling is given by three peaks, while as $N$ is increased the distribution is driven to consist of two peaks.

Let us now discuss the solution of the maximum entropy problem within the two block ansatz: the magnetization in block A is

$$m_A = \frac{1}{N_A} \sum_{i=1}^{N_A} \sigma_i, \qquad \text{(S8)}$$

and similarly for $m_B$. By using the block ansatz for the correlation matrix $C_{ij}$ shown in Fig. S3, we can write Eq. (7) as

$$P(\sigma) = \frac{1}{Z} \exp(N \vec{m}^T \cdot \mathcal{M} \cdot \vec{m}), \qquad \text{(S9)}$$

where $\vec{m} \equiv (m_A, m_B)$ and $\mathcal{M}$ is given by Eq. (12). The sum over spin configurations can be written as an integral over the order parameters $m_A$ and $m_B$ by standard

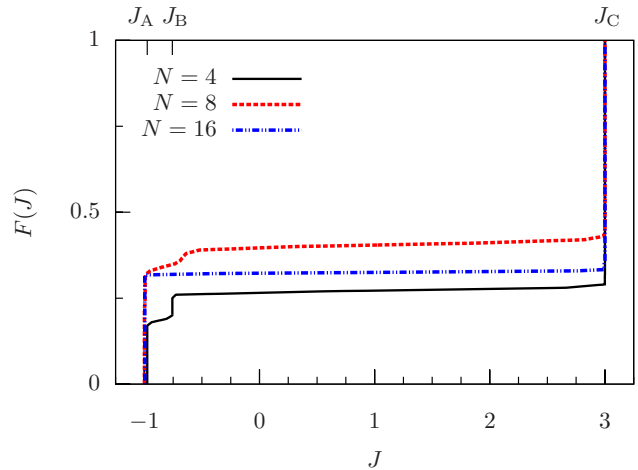

FIG. S2: Cumulative distribution function of couplings for coupled spins in the two moment case. For a given $N$ and $\lambda_1 = \lambda_2 = 1$, Eq. (8) is solved numerically multiple times starting from different random initial conditions for $\{J_{ij}\}$: as a result, a total number $512 \leq S \leq 65536$ of couplings $J_{ij}$ is obtained from this ensemble of solutions. The cumulative distribution function $F(J)$ is defined as the number of couplings $J_{ij}$ smaller than $J$ divided by the total number of couplings $S$, and it is plotted as a function of $J$ for system sizes $N = 4, 8, 16$. For $N = 4$, the $J_{ij}$s cluster tightly around three values $J_A, J_B, J_C$, and by $N = 16$ the distribution of $J_{ij}$s is almost perfectly discrete, with only two values $J_A, J_C$.

$$C_{ij} = \left( \begin{array}{|c|c|} \hline C_I & C_{II} \\ \hline C_{II} & C_I \\ \hline \end{array} \right) \begin{array}{l} \} N_A \\ \} N_B \end{array}$$
$$\underbrace{\phantom{xxxxxxxxxx}}_{N}$$

FIG. S3: Block ansatz for the correlation matrix $C_{ij}$: correlations between spins within a block are equal to $C_I$, while correlations between spins in different blocks are equal to $C_{II}$.

methods. The result for the partition function is

$$Z = \frac{2\pi N}{\sqrt{\det \mathcal{M}}} \int d\vec{m} \, \exp\left[-N\mathcal{S}(\vec{m})\right], \qquad \text{(S10)}$$

where $\mathcal{S}$ is given by Eq. (13). The integral in Eq. (S10) is dominated by the saddle point $\vec{m}^*$ which minimizes $\mathcal{S}$: let us consider first the high temperature case, where we match the observed correlations with a model with saddle point $\vec{m}^* = \vec{0}$. The free energy is

$$f = -\ln 2 + \frac{1}{2N} \ln \mathscr{D}, \qquad \text{(S11)}$$

where $\mathscr{D} = \det\left(1 - 2\,\mathrm{diag}(x, 1-x)^{-1} \cdot \mathcal{M}\right)$. The entropy per spin is given by Eq. (S6), and Eq. (11) is then obtained from Eqs. (S6,S11). The entropy (S6) can be written as the difference between an energy term and the free energy. Hence, the maximum of the entropy is obtained as a tradeoff between free energy and energy.

This maximum is reminiscent of the minimum of the free energy in the direct problem, which is obtained as a tradeoff between energy and entropy.

In the low temperature case, the saddle point is at $\vec{m} = \vec{m}^* \neq \vec{0}$. The free energy is given by $f = \mathcal{S}(\vec{m}^*)$, and it can be computed numerically for any given $\lambda_1, \lambda_2, x$: the optimal value of $x$ can be obtained along the same lines as in the high temperature case.

**Three moment case**

Here we will show that our maximum entropy solution reproduces an Ising spin glass if we constrain three moments of the correlations. The main feature of an Ising spin glass is the presence of frustration: namely, the product of the interactions $J_{ij}$ along a loop of spins has a negative sign. In order to obtain a frustrated maximum entropy solution, one needs to split the coupling matrix $J_{ij}$ into at least three blocks. As a consequence, a spin-glass maximum entropy solution can be obtained only if at least three moments of the correlations are constrained. This can be done along the lines of the two moment case, and we will sketch the main steps here. We maximize the entropy (S6) by constraining the first three moments $C_1, C_2, C_3$ of the correlations: the Boltzmann distribution is still given by Eq. (7), with

$$J_{ij} = \lambda_1 + 2\lambda_2 C_{ij} + 3\lambda_3 C_{ij}^2. \quad (S12)$$

We construct a three block correlation matrix $C_{ij}$ starting from the two block case: we consider the bottom-right block in Fig. S3, and we partition it into blocks as we did for the matrix $C_{ij}$ in the two moment case. As a result, spins are now divided into groups $A \equiv \{\sigma_1, \cdots, \sigma_{N_A}\}$, $B \equiv \{\sigma_{N_A+1}, \cdots, \sigma_{N_A+N_B}\}$ and $C \equiv \{\sigma_{N_A+N_B+1}, \cdots, \sigma_N\}$, and correlations $C_{ij}$ take three values $C_I, C_{II}, C_{III}$. In the high temperature phase where $C_1 = \mathcal{O}(1/N)$, $C_2 = \mathcal{O}(1/N^2)$, $C_3 = \mathcal{O}(1/N^3)$, the free energy is given by Eq. (S11), with $\mathscr{D} = \det\left(1 - 2\,\mathrm{diag}(x, y, 1-x-y)^{-1} \cdot \mathscr{M}\right)$ where $x = N_A/N$, $y = N_B/N$, and the matrix $\mathscr{M}$ is given by

$$\mathscr{M} = \begin{pmatrix} x^2(\lambda_1 + 2\lambda_2 C_I + 3\lambda_3 C_I^2) & x(y\lambda_1 + 2u\lambda_2 C_{II} + 3u\lambda_3 C_{II}^2) & x(z\lambda_1 + 2u\lambda_2 C_{II} + 3u\lambda_3 C_{II}^2) \\ x(y\lambda_1 + 2u\lambda_2 C_{II} + 3u\lambda_3 C_{II}^2) & y^2(\lambda_1 + 2\lambda_2 C_I + 3\lambda_3 C_I^2) & yz(\lambda_1 + 2\lambda_2 C_{III} + 3\lambda_3 C_{III}^2) \\ x(z\lambda_1 + 2u\lambda_2 C_{II} + 3u\lambda_3 C_{II}^2) & yz(\lambda_1 + 2\lambda_2 C_{III} + 3\lambda_3 C_{III}^2) & z^2(\lambda_1 + 2\lambda_2 C_I + 3\lambda_3 C_I^2) \end{pmatrix}, \quad (S13)$$

with $z = 1 - x - y$, $u = 1 - x$. Given $C_1, C_2, C_3$ and $x, y$, we determine $C_I, C_{II}, C_{III}$ and $\lambda_1, \lambda_2, \lambda_3$ by matching the moments of correlations as in Eq. (9), and then we maximize $f$ with respect to $x, y$.

In an Ising spin glass we expect spin-spin correlations to have both positive and negative signs: hence, the odd moments of correlations are small compared to the even ones. Indeed, if the correlation moments satisfy these conditions, we obtain that our maximum entropy solution involves frustrated spin-spin interactions typical of a spin glass. Explicitly, we take $C_1 = 0.1$, $C_2 = 0.5$, $C_3 = 0.05$, and we obtain that the interactions along a loop connecting spins in blocks A, B, C are $J_{AB} = -0.13$, $J_{BC} = -0.098$, $J_{CA} = J_{AB}$, implying that $J_{AB} J_{BC} J_{CA} < 0$.

As we fix more and more moments, we expect our maximum entropy method to reproduce the full features of a spin glass, such as the emergence of a critical behavior. To identify the scaling of correlations corresponding to the critical regime, we recall that given the overlap $Q \equiv \frac{1}{N} \sum_{i=1}^N \sigma_i^1 \sigma_i^2$ between two independent replicas $\sigma^1, \sigma^2$, in the high temperature regime we have $\langle Q^2 \rangle = \mathcal{O}(1/N)$ [1], hence for large $N$

$$\begin{aligned} C_2 &= \frac{1}{N_p} \sum_{i>j} \langle \sigma_i \sigma_j \rangle^2 \quad (S14) \\ &= \frac{1}{(1 - 1/N)} \left( \frac{1}{N^2} \sum_{ij} \langle \sigma_i \sigma_j \rangle^2 \right) - \frac{1}{N} \\ &= \langle Q^2 \rangle - \frac{1}{N}. \end{aligned}$$

It follows that the critical spin glass regime is obtained as $C_1 \approx C_3 \approx 0$, $C_2 \gg \mathcal{O}(1/N)$.



\end{document}